\documentclass[12pt]{article}
\usepackage{setspace}
\pdfoutput = 1
\usepackage[utf8]{inputenc}
\usepackage{dsfont}
\usepackage{epsfig}
\usepackage{amsmath}
\usepackage{amssymb}
\usepackage{stmaryrd}
\usepackage{empheq}
\usepackage{esint}
\usepackage{physics}
\usepackage{bbm}
\usepackage{amsfonts}
\usepackage{array}
\usepackage{setspace}
\usepackage{braket}
\usepackage{float}
\usepackage{url}
\usepackage{mathtools}
\usepackage{tensor}
\usepackage{bbold}
\usepackage{slashed}
\usepackage{tikz}
\usepackage{svg}
\allowdisplaybreaks
\usepackage{changepage}

\usepackage[margin = 1in]{geometry}
\usepackage[ragged]{footmisc}
\usepackage[bookmarks=true,bookmarksnumbered=false,
hyperindex=true,bookmarksopen=true,hyperfigures=true,
colorlinks=true,linkcolor=webblue,citecolor=webgreen,urlcolor=webblue,breaklinks]{hyperref}
\definecolor{webgreen}{rgb}{0, 0.5, 0}
\definecolor{webblue}{rgb}{0, 0, 0.5}
\definecolor{webred}{rgb}{0.5, 0, 0}
\definecolor{darkgreen}{rgb}{0,0.5,0}

\begin{document}

\thispagestyle{empty}
 ~\vspace{5mm}
\begin{adjustwidth}{-1cm}{-1cm}
\begin{center}
 {\Large \bf 
The Observer Paradigm in Semiclassical Gravity: A Postmodern Perspective}
 \vspace{0.4in}

 {Sean McBride}
 \end{center}
 \end{adjustwidth}

\begin{center}
{March 31, 2026}

 \vspace{0.1in}
 {Department of Physics and Astronomy, University of British Columbia \\ 6224 Agricultural Road, Vancouver, BC V6T 1Z1}
 \vspace{0.1in}
 
 {\tt samcbride@phas.ubc.ca}
\end{center}


\begin{abstract}
\noindent
Recent advances in semiclassical gravity, both in our understanding of the gravitational path integral and the algebraic structure of a gravitating subregion, rely on the presence of an observer to obtain a nontrivial Hilbert space for closed universe backgrounds. Here I examine this proposal from a postmodern lens, identifying attempts to define ``observer rules'' as manifestations of metanarrative breakdown: the observer both supplies and undermines the perturbative gravitational Hilbert space. Rather than resolving this tension, I advocate for a post-postmodern acceptance of the incompatibility between observer-dependent and observer-free descriptions of closed universes, treating the ambiguity as a feature of quantum gravity's fundamental degrees of freedom. To my knowledge, this is the first reference to post-postmodernism in concert with physics.
\end{abstract}
\vspace{1 in}
\begin{center} \textit{Dedicated to the memory of Timothy Charles McBride (1966-2026); Bravo Zulu}

\vspace{1 in}
{\small Essay written for the Gravity Research Foundation 2026 Awards for Essays on Gravitation}
\end{center}

\pagebreak
\setcounter{page}{1}
\setcounter{tocdepth}{2}

\section*{Introduction}

The status of the observer in quantum mechanics is one of detachment and omnipotence. In a practical setting, the observer is the experimenter: turning dials, choosing inputs, and recording outputs. This p-zombie (the ``p'' standing for physics) is not regularly discussed in quantum mechanics nor in quantum field theory; the usual treatment is to fully abstract the observer from the scenario, replacing them with a discussion of measurement. The former notion of an observer, an omnipresent being conducting an experiment, is eradicated and replaced by a series of mathematically well-defined but emotionally obtuse positive operator-valued measures (POVMs) and the now century-old Born rule \cite{Born1926-et}. This is par for the course in high-energy physics, where the outputs of a putative physical theory are observables: quantities computed via a functional integral used as a stopgap to model the experience of an observer.

Going back to Wheeler \cite{Wheeler1978-qy, WheelerLaw, Wheeler1989-WHEIPQ}, gravity theorists have questioned what changes when the observer is observing the universe they themselves are part and parcel of. These questions became even more relevant when considering the black hole information paradox, where the original question of whether information is able to escape morphed into the question of detection of a Page curve by an outside experimenter, redirecting attention from the observed to the observer \cite{Susskind:1993if, Almheiri:2012rt, Harlow:2013tf}. One striking output from this direction was verification of a principle only known as lore: closed universes in quantum gravity have trivial (i.e. one-dimensional) Hilbert spaces! There are many avenues to this conclusion: the gauging of time translation, the absence of a holographically-relevant physical boundary, the reality of the closed universe Hilbert space under gauging spacetime inversion symmetries \cite{Harlow:2023hjb} just to name a few. To address this issue, attention shifted from discussing a Hilbert space, a difficult concept to define for field theories, to defining an algebra of observables, a set of operations on a Hilbert space which only makes use of a putative vacuum state.

But who is performing these operations? In the case of a subregion in de Sitter, this algebra is only defined nontrivially in the case where one appends an observer, modelled by an observer Hilbert space $\mathcal{H}_{\text{obs}} = L^2(\mathbbm{R}_+)$ \cite{Chandrasekaran:2022cip}.\footnote{The choice of $\mathbbm{R}_+$ is meant to emulate a measuring device with a ground state and a continuum of measurable states.} The inclusion of an observer spurred a study of the necessity of observers for closed universe dynamics, coterminous with a quest to define ``observer rules'': guidelines on how to treat the piece of the puzzle outside the typical rules inherited from the gravitational path integral. The proposals for the observer rules are myriad, though their mathematical conclusions are fairly universal, with the ontological status of the observer still up in the air. My goal in this essay is to view these various proposals using the precepts of postmodernism, arguing that the observer can be concretely interpreted as a narrator which is both integral and detrimental to the emergent gravitational Hilbert space. I'll maintain a laser focus on the observer paradigm, leaving broader questions in the philosophy of science to the philosophers. The details of the individual proposals will not be too relevant, as I'll prioritize their universalities and overall mission statements as opposed to the technical details, which hopefully will serve as an olive branch to those not in the know. 

\section{A Postmodern Primer}

Before seriously evoking postmodernism, one must define postmodernism, and before defining postmodernism, one must define modernism.\footnote{A time-reversed ``If You Give a Physicist a Philosophy Wiki'', if you will.} Both concepts are best explained in a historical context. Modernism, broadly defined, was meant to shake off the shackles of stodgy tradition in favor of progress, perhaps best stated as the existence of a grand societal narrative (hereafter the metanarrative), of which science was and is an excellent provider. By viewing the progress of science as a benchmark for the progress of society, modernist thinking abounded in the wake of the industrial revolution. 

Postmodernism was, in many ways, a direct result of cynicism in postwar Europe (referring both to World Wars \cite{toynbee1939study}) and the counter-cultural movement of the 1960s. Its proponents go a step further, not only rejecting tradition as a supplier of the metanarrative, but rejecting the very concept of a reliable metanarrative, instead resorting to skepticism and cultural critique as a dominant form of interaction with the world. Both modernism and postmodernism have permeated all aspects of the humanistic world including art, architecture, literature, and film, but our focus will be on science, the \textit{sine qua non} of modernist thinking.

\subsection{Postmodern Physics}

The use of the term ``paradigm'' in the title of this essay is deliberate, meant to evoke the Kuhnian notion of paradigm shifts \cite{kuhn1970structure}. Kuhn's foundational text, \textit{The Structure of Scientific Revolutions}, identifies two kinds of science: normal science and revolutionary science, the latter of which ushers in paradigm shifts, the umbrellas under which normal science is performed on the daily. The final say on what constitutes a paradigm, whether it's heliocentrism or relativity or the existence of oxygen, is a complex social process, as it's only due to the subjectivity of individual scientists that an idea becomes accepted at large.

Postmodern discourse in science follows directly from Kuhn's study of the interpersonal politics of science, though not necessarily from his subsequent dichotomization of the field.\footnote{A notion which Kuhn distanced himself from not long after \textit{SSR}'s publication and which hasn't really caught on \cite{Bird2004}.} A putative manifesto for postmodernism can be found in Lyotard \cite{Lyotard1984-mm}, where the Kuhnian perspective on scientific knowledge is extended and subsumed into Lyotard's concept of metanarrative breakdown, characterized by incredulity and skepticism towards the influence of fallible systems of power. The scientific analogue of a metanarrative is the positivist notion of linear scientific progress.

After Lyotard's influential work, postmodernist thought and physics enjoyed a brief fling. As was well explicated by Carson \cite{Carson1995-od}, postmodern physics of that era was first and foremost concerned with what might be termed the unreliability of physical paradigms. Of particular focus were the deterministic but practically unpredictable chaos theory and the probabilistic nature of the measurement problem in quantum mechanics. These concepts, though, were communicated through a process Carson terms ``philosophical popularization'', the unavoidable domination of postmodern physics by what popular physics books were in the zeitgeist and digestible by a largely orthogonal community. After the science wars of the late 1990s, postmodernism (and moreover critical theory in general) was often equated with postempiricism, and by proxy anti-intellectualism \cite{Sokal1999-br, Smolin2006-is}, and references to postmodernism and physics dwindled. 

It's important to note, in light of this warped view, that the postmodern perspective does not deny the existence of a metanarrative; rather, it questions the legitimacy of such narratives and critically examines their sources. In that way, it should be viewed neither as anti-empirical, nihilistic, nor devoid of intellectual meaning, as was previously espoused. I do feel, however, that orthodox physics, vouchsafed by centuries of experiments and fine-tuning, is by and large inscrutable to a postmodern viewpoint. The goal of this essay is to display a new brand of postmodernism in physics, one that features a physically-informed perspective only relevant for a small class of questions in semiclassical gravity. 

\section{The Observer Paradigm}

There is, as of writing this essay, spirited debate about the observer rules, that is how to treat the observer with the same philosophy with which one treats the gravitational path integral, which at least in some frameworks can be treated axiomatically at the level of rigor physicists will accept. Here I review the salient points.

The observer paradigm involves finding an auxiliary observer Hilbert space. This Hilbert space can be put in by hand, as is the case of the de Sitter construction \cite{Chandrasekaran:2022cip} or quantum circuit models \cite{Harlow:2025pvj, Engelhardt:2025azi, Akers:2025ahe}, or can be built from degrees of freedom in the semiclassical bulk field theory \cite{Abdalla:2025gzn, Antonini:2025ioh, Chen:2024rpx}. It's crucial that, ignoring certain non-perturbative effects, the observer Hilbert space tensor factorizes from the remaining bulk degrees of freedom; this can be accomplished by relaxing the gravitational constraints, modifying the non-isometric code mapping to the fundamental degrees of freedom, or altering the rules of the sum over topologies in the gravitational path integral.

Despite their differences, these constructions lead to broadly similar conclusions. Via a clever application of these observer rules, one finds a nontrivial Hilbert space, one which depends not only on the gravitational entropy but also on the intrinsic entropy of the observer. As a concrete example, in the HUZ model \cite{Harlow:2025pvj} applied to JT gravity, the fundamental closed universe Hilbert space is approximately
\begin{equation}
    \text{dim } \mathcal{H}_{\text{fun}} \approx \min \left( e^{S_{\text{obs}}}, e^{2S_0} \right),
    \label{eq1}
\end{equation}
whereas in the AAIL model \cite{Abdalla:2025gzn} in the same theory, one finds
\begin{equation}
    \text{dim } \mathcal{H}_{\text{fun}} \approx e^{S_{\text{obs}} + 2S_0},
    \label{eq2}
\end{equation}
for an observer entropy $S_{\text{obs}}$ and $S_0$ is the extremal entropy analogous to $1/G_N$. The observer is therefore the supplier of dynamics in the closed universe! Moreover, the inclusion of an observer was recently used to clarify a confusion in the state-counting interpretation of Euclidean de Sitter space \cite{Maldacena:2024spf, Chen:2025jqm}, again demonstrating the inseparability of the observer from the observable.

The notion that an observer is limited by their intrinsic entropy is not in and of itself surprising, but it does impose a constraint on the observer typically glossed over in standard quantum mechanics.\footnote{It has been said that quantum mechanics is an effective theory in large $N$ where $N$ is the number of atoms in the brain. I first heard this comment from Daniel Harlow.} What is shocking is that the Hilbert space accessible by the observer to an acceptable degree of accuracy is itself dependent on the size of the observer, as can be seen in equations \eqref{eq1} and \eqref{eq2}. This differs from standard breakdowns of effective field theory in semiclassical gravity, e.g. the statement that correlation functions of single-trace operators in AdS/CFT fail to make sense on a fixed background when the number of operators scales with $N^2$. Whereas limits on numbers of operators or scaling dimensions in correlation functions are provided by fundamental constants in the effective theory, the observer paradigm instead limits the accessible Hilbert space via a seemingly ad hoc observer entropy. This is a true ceiling on the observer's experimental capabilities, one they're unable to shake as they are confined by their boundaryless spacetime.

\begin{figure}
    \centering
    \includegraphics[width=0.7\linewidth]{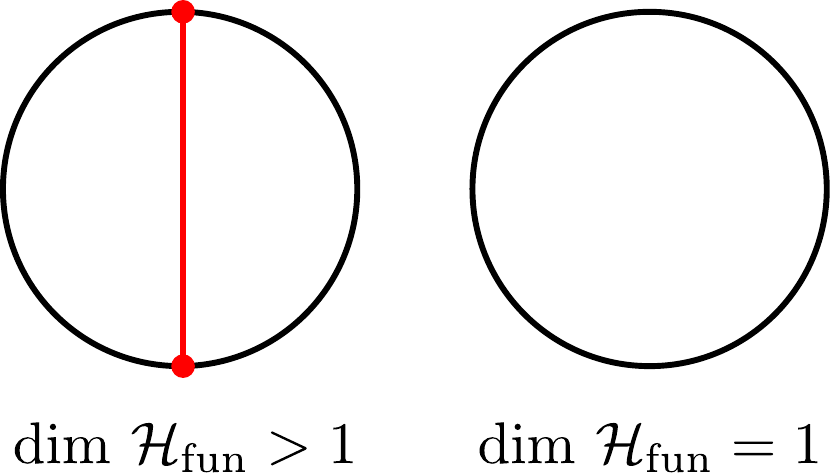}
    \caption{The two fundamental Hilbert spaces that result from the inclusion or exclusion of an observer. The Hilbert space in the observer-laden closed universe is an approximate description capped off by the observer's intrinsic entropy. The Hilbert space of the observer-free closed universe is nonperturbatively trivial.}
    \label{fig:fig1}
\end{figure}

\subsection{The Observer-Narrator}

I'd like to argue that these various attempts at defining observer rules, and their perhaps unwelcome consequences, emphasize precisely the tenets of postmodernism, namely the breakdown of the na\"ively supplied metanarrative. 

The inclusion of an observer is indeed unnatural from the Wilsonian pathway high-energy physics has operated under for the better part of a century \cite{Wilson:1971bg, Wilson:1971dh}. In particle physics, new degrees of freedom are meant to solve a problem and reduce complexity, but here the observer's unclear origin raises more questions than it answers. Given that the observer is a requirement to continue doing physics as we're accustomed to, the observer assumes the role of a totalitarian narrator, the metanarrative its reflexively-defined fundamental Hilbert space.

Due to this abrupt shift in power from the theorist at the keyboard to the unabstracted observer, the questions under consideration now are of a completely different flavor. Instead of asking about unification, the observer rules are concerned with separation, poking and prodding the privileged observer who so generously provided the physics in the first place. Is it correct to model the observer Hilbert space by $L^2(\mathbb{R}_+)$, under the simple principle that measured energies should be bounded by below? Should the observer worldline remain topologically distinct from matter worldlines in JT gravity? Must the observer be put in by hand or should they be apparent from the low energy effective theory? The functional discrepancy between equations \eqref{eq1} and \eqref{eq2} is an avatar of this ontological confusion.

Moreover, the inability to remove the observer-narrator and retain a nontrivial Hilbert space calls into question whether this is a truly fundamental theory of quantum gravity, which is often thought to have no free parameters. The unwelcome inclusion of choice makes it tempting to consider the observer as emergent, just as we view spacetime itself as emergent \cite{VanRaamsdonk:2010pw}. Unfortunately, emergence from a one-dimensional Hilbert space is ill-defined. Whereas one would like a gravitational Hilbert space whose dimension depends solely on gravitational constants, it was found to depend sensitively on, and in fact be bottlenecked by, the observer's intrinsic Hilbert space. The observer-narrator, the supplier of nontrivial dynamics in the closed universe, is both the prisoner and the jailer. 

Again I should stress that \textit{any} observer in quantum mechanics is experimentally limited by their intrinsic entropy. However, outside of the setting of an observer in a closed universe, this inconvenience is irrelevant due to the nonperturbative extradition of the experimenter from the subsystem under study. The observer paradigm's inevitable self-reference necessitates postmodern thought, again and again shifting our focus to epistemological matters, rather than leaving us content with our standard, empirically-informed physical motivations.

\section{A Post-Postmodern Physics?}

An obligatory line of inductive thought leads us to ask this provocative question.\footnote{I'm grateful to Jeremy van der Heijden, Marc Klinger, and Alejandro Vilar L\'opez for a fruitful discussion which helped me clarify my position in this section.} Post-postmodernism\footnote{The comments in this section apply equally well to New Sincerity, metamodernism, post-irony and the like.} is roughly what it sounds like: a reaction to postmodernism, just as postmodernism was a reaction to modernism. Rather than emphasizing the abstraction from metanarratives inherent in postmodern thinking, it embraces the abstraction itself as a source of knowledge. In some ways, post-postmodernism closes the distance enforced by postmodernism, opening up avenues for sincere thinking without the need to consider the systems of power which restrict that thinking, as postmodernism's irony is itself providing a potentially unwelcome and power-enforcing metanarrative. To deescalate the self-absorbed language for a moment, post-postmodernism is the recognition that some problems don't need to be philosophically complicated, they can just be.

The application of this attitude towards physics is difficult to unpack. After all, post-postmodernism is predominantly concerned with our interaction with popular culture, media, technology, namely the post-ironic monoculture advocated for in e.g. \cite{DFW}. Attempting to define a monoculture of physics in the framework we've put forth would be beyond the scope of this, or indeed any, work. The moving-beyond in the case of the observer paradigm, in my view, is precisely the acceptance of the disparity between the observer paradigm and the mandates of the gravitational path integral. A constant struggle for those in this framework is the acceptance of a one-dimensional Hilbert space for a closed universe. Does this mean our universe has a boundary? How can nontrivial dynamics emerge from a trivial Hilbert space? And if dynamics can't emerge, how can we make it emerge? Is it really fundamental to have two ``fundamental'' theories?

A post-postmodern perspective, and indeed my own perspective, is to instead accept the breakdown of predictability and fundamentality inherent in the observer paradigm, as well as the triviality of the observer-free closed universe Hilbert space. I find  incompatibility in including an observer for questions involving measurement, experiments, and what would ordinarily be termed physics while honoring a formalism which returns a trivial Hilbert space for an observer-free closed universe. After all, the universe I inhabit has at least one nonnegotiable observer --- me! This is not throwing in the towel; rather it is a sincere recognition that the tension between the two descriptions is constitutive of what quantum gravity in a closed universe actually is, instead of a symptom of our incomplete understanding of it. In this sense the post-postmodern move is a positive characterization instead of a demand for surrender. If the only objections to a one-dimensional closed universe are couched in an empirical or phenomenological language, then I believe calling the observer-laden Hilbert space fundamental is a valid choice. If they arise from an epistemological complaint about the notion of the fundamental, then we'll have to call it a draw.

Putting all of this together reinforces my viewpoint that there is no need for the observer to ``emerge'' from the nonexistent degrees of freedom of a trivial closed spacetime; rather, acceptance of the incompatibility of the two descriptions seems like a perfectly cromulent piece of the definition of the fundamental degrees of freedom in quantum gravity.

\section*{Summary}

Titling this section \textbf{Conclusions} would be an act of war, as I do not deign to have made scientific conclusions in any theory reasonably termed scientific. Here I just offer a recapitulation of the main themes.

Postmodern thought is couched in the belief that truth is inextricable from the systems of power that generate it, a belief which inevitably leads to skepticism and irony as dominant modalities of interaction with the world. Rather than focus on the implications of postmodern knowledge and thought in science, I choose instead to concentrate on a small corner of that knowledge, namely the observer paradigm in semiclassical gravity. This is an avenue where the value of a physical concept, the observer, manifestly holds narrative power over physical principles, power which self-sabotages experiments performed by that very observer. The observer introduces an ambiguity in the fundamental description of closed universes in quantum gravity, an ambiguity I don't resolve but instead accept, in a sense denying the observer-narrator's power of obfuscation.

The inquiries of those working in the observer paradigm, and more generally the It from Qubit program, are not at all correlated with empirical observations. The decline of experimental interaction is a necessary evil; certain questions we as a scientifically advanced society want to answer are inaccessible to the most highfalutin of laboratory work. It's worth reiterating that framing the problem of the observer as a development in postmodern thought is not meant as ridicule nor a comment on the post-empiricist nature of contemporary high-energy physics. Rather, this work is perhaps best read as a new thread in a torn web, weaving together mandates from the gravitational path integral with developments in critical theory necessitated by a sufficiently detached perspective on knowledge.

The observer rules, as they have been so coined, are partially inserted by fiat, under reasonable assumptions of what should constitute an observer and their lived experience performing sufficiently localized experiments. A criticism of this approach is the addition of choice; in contrast, the inclusion of replica wormholes in the gravitational path integral to derive the Page curve is not, at first light, a radical departure from the mandates of the path integral in quantum field theory. I've argued that the choices of the observer paradigm are not a failure of high energy physicists to uncover the full Theory of Everything; they should be taken as an honest reckoning with the fact we are stuck inside the universe we're trying to describe.  

\noindent\rule{\textwidth}{1pt}

\begin{quote}
    \textit{The cosmologies of today---big bangs and black holes, antimatter and curved, ever-expanding space going nowhere---leave us in dread and seeming incomprehensibility. Random events, nothing truly necessary. Science's cosmologies say nothing about the soul, about its reason for existence, how it comes to be and where it might be going, and what its tasks could be. \cite{Hillman1997-zd}}
\end{quote}

\subsection*{Acknowledgments}

I would like to acknowledge Thomas Frazier, David Grabovsky, Jeremy van der Heijden, Marc Klinger, Michael Sipiora, and Alejandro Vilar L\'opez for productive discussions. I'd also like to thank Katherine Lin for accommodation where this work was initiated. I am supported by the National Science and Engineering Council of Canada (NSERC) and the Simons Foundation via a Simons Investigator Award.

\bibliographystyle{JHEP.bst}
\bibliography{main2}

\providecommand{\href}[2]{#2}\begingroup\raggedright\begin{thebibliography}{10}

\bibitem{Born1926-et}
M.~Born, \emph{Zur quantenmechanik der stoßvorgänge}, {\emph{Eur. Phys. J. A} {\bfseries 37} (1926) 863}.

\bibitem{Wheeler1978-qy}
J.A.~Wheeler, \emph{The ``past'' and the ``delayed-choice'' double-slit experiment},  in \emph{Mathematical Foundations of Quantum Theory}, pp.~9--48, Elsevier (1978).

\bibitem{WheelerLaw}
J.A.~Wheeler, \emph{Law without law},  in \emph{Quantum Theory and Measurement}, pp.~182--213, Princeton University Press (1983).

\bibitem{Wheeler1989-WHEIPQ}
J.A.~Wheeler, \emph{Information, physics, quantum: The search for links},  in \emph{Proceedings III International Symposium on Foundations of Quantum Mechanics}, W.J.~Archibald, ed., pp.~354--358 (1989).

\bibitem{Susskind:1993if}
L.~Susskind, L.~Thorlacius and J.~Uglum, \emph{{The Stretched horizon and black hole complementarity}}, \href{https://doi.org/10.1103/PhysRevD.48.3743}{\emph{Phys. Rev. D} {\bfseries 48} (1993) 3743} [\href{https://arxiv.org/abs/hep-th/9306069}{{\ttfamily hep-th/9306069}}].

\bibitem{Almheiri:2012rt}
A.~Almheiri, D.~Marolf, J.~Polchinski and J.~Sully, \emph{{Black Holes: Complementarity or Firewalls?}}, \href{https://doi.org/10.1007/JHEP02(2013)062}{\emph{JHEP} {\bfseries 02} (2013) 062} [\href{https://arxiv.org/abs/1207.3123}{{\ttfamily 1207.3123}}].

\bibitem{Harlow:2013tf}
D.~Harlow and P.~Hayden, \emph{{Quantum Computation vs. Firewalls}}, \href{https://doi.org/10.1007/JHEP06(2013)085}{\emph{JHEP} {\bfseries 06} (2013) 085} [\href{https://arxiv.org/abs/1301.4504}{{\ttfamily 1301.4504}}].

\bibitem{Harlow:2023hjb}
D.~Harlow and T.~Numasawa, \emph{{Gauging spacetime inversions in quantum gravity}}, \href{https://doi.org/10.1007/JHEP01(2026)098}{\emph{JHEP} {\bfseries 01} (2026) 098} [\href{https://arxiv.org/abs/2311.09978}{{\ttfamily 2311.09978}}].

\bibitem{Chandrasekaran:2022cip}
V.~Chandrasekaran, R.~Longo, G.~Penington and E.~Witten, \emph{{An algebra of observables for de Sitter space}}, \href{https://doi.org/10.1007/JHEP02(2023)082}{\emph{JHEP} {\bfseries 02} (2023) 082} [\href{https://arxiv.org/abs/2206.10780}{{\ttfamily 2206.10780}}].

\bibitem{toynbee1939study}
A.J.~Toynbee, \emph{A Study of History}, vol.~5, Oxford University Press, London (1939).

\bibitem{kuhn1970structure}
T.S.~Kuhn, \emph{The Structure of Scientific Revolutions}, University of Chicago Press, Chicago, 2~ed. (1970).

\bibitem{Bird2004}
A.~Bird, \emph{Kuhn and twentieth century philosophy of science}, \href{https://doi.org/10.4288/jafpos1956.12.61}{\emph{Annals of the Japan Association for Philosophy of Science} {\bfseries 12} (2004) 1}.

\bibitem{Lyotard1984-mm}
J.-F.~Lyotard, \emph{The Postmodern Condition: A Report on Knowledge}, Theory and History of Literature, University of Minnesota Press, Minneapolis, MN (Mar., 1984).

\bibitem{Carson1995-od}
C.~Carson, \emph{Who wants a postmodern physics?}, {\emph{Sci. Context} {\bfseries 8} (1995) 635}.

\bibitem{Sokal1999-br}
A.~Sokal and J.~Bricmont, \emph{Fashionable Nonsense: Postmodern Intellectuals' Abuse of Science}, Picador, New York, NY (Nov., 1999).

\bibitem{Smolin2006-is}
L.~Smolin, \emph{The Trouble with Physics: The Rise of String Theory, the Fall of a Science, and What Comes Next}, Houghton Mifflin (Sept., 2006).

\bibitem{Harlow:2025pvj}
D.~Harlow, M.~Usatyuk and Y.~Zhao, \emph{{Quantum mechanics and observers for gravity in a closed universe}}, \href{https://doi.org/10.1007/JHEP02(2026)108}{\emph{JHEP} {\bfseries 02} (2026) 108} [\href{https://arxiv.org/abs/2501.02359}{{\ttfamily 2501.02359}}].

\bibitem{Engelhardt:2025azi}
N.~Engelhardt, E.~Gesteau and D.~Harlow, \emph{{Observer complementarity for black holes and holography}},  \href{https://arxiv.org/abs/2507.06046}{{\ttfamily 2507.06046}}.

\bibitem{Akers:2025ahe}
C.~Akers, G.~Bueller, O.~DeWolfe, K.~Higginbotham, J.~Reinking and R.~Rodriguez, \emph{{On observers in holographic maps}}, \href{https://doi.org/10.1007/JHEP05(2025)201}{\emph{JHEP} {\bfseries 05} (2025) 201} [\href{https://arxiv.org/abs/2503.09681}{{\ttfamily 2503.09681}}].

\bibitem{Abdalla:2025gzn}
A.I.~Abdalla, S.~Antonini, L.V.~Iliesiu and A.~Levine, \emph{{The gravitational path integral from an observer{\textquoteright}s point of view}}, \href{https://doi.org/10.1007/JHEP05(2025)059}{\emph{JHEP} {\bfseries 05} (2025) 059} [\href{https://arxiv.org/abs/2501.02632}{{\ttfamily 2501.02632}}].

\bibitem{Antonini:2025ioh}
S.~Antonini, P.~Rath, M.~Sasieta, B.~Swingle and A.~Vilar~L{\'o}pez, \emph{{The baby universe is fine and the CFT knows it: on holography for closed universes}}, \href{https://doi.org/10.1007/JHEP12(2025)159}{\emph{JHEP} {\bfseries 12} (2025) 159} [\href{https://arxiv.org/abs/2507.10649}{{\ttfamily 2507.10649}}].

\bibitem{Chen:2024rpx}
C.-H.~Chen and G.~Penington, \emph{{A clock is just a way to tell the time: gravitational algebras in cosmological spacetimes}},  \href{https://arxiv.org/abs/2406.02116}{{\ttfamily 2406.02116}}.

\bibitem{Maldacena:2024spf}
J.~Maldacena, \emph{{Real observers solving imaginary problems}},  \href{https://arxiv.org/abs/2412.14014}{{\ttfamily 2412.14014}}.

\bibitem{Chen:2025jqm}
Y.~Chen, D.~Stanford, H.~Tang and Z.~Yang, \emph{{On the phase of the de Sitter density of states}},  \href{https://arxiv.org/abs/2511.01400}{{\ttfamily 2511.01400}}.

\bibitem{Wilson:1971bg}
K.G.~Wilson, \emph{{Renormalization group and critical phenomena. 1. Renormalization group and the Kadanoff scaling picture}}, \href{https://doi.org/10.1103/PhysRevB.4.3174}{\emph{Phys. Rev. B} {\bfseries 4} (1971) 3174}.

\bibitem{Wilson:1971dh}
K.G.~Wilson, \emph{{Renormalization group and critical phenomena. 2. Phase space cell analysis of critical behavior}}, \href{https://doi.org/10.1103/PhysRevB.4.3184}{\emph{Phys. Rev. B} {\bfseries 4} (1971) 3184}.

\bibitem{VanRaamsdonk:2010pw}
M.~Van~Raamsdonk, \emph{{Building up spacetime with quantum entanglement}}, \href{https://doi.org/10.1142/S0218271810018529}{\emph{Gen. Rel. Grav.} {\bfseries 42} (2010) 2323} [\href{https://arxiv.org/abs/1005.3035}{{\ttfamily 1005.3035}}].

\bibitem{DFW}
D.F.~Wallace, \emph{E unibus pluram: Television and u.s. fiction}, {\emph{Review of Contemporary Fiction} {\bfseries 13:2} (1993) 151}.

\bibitem{Hillman1997-zd}
J.~Hillman, \emph{The Soul's Code: In Search of Character and Calling}, Warner Books, New York, NY (Oct., 1997).

\end{thebibliography}\endgroup

\end{document}